\documentclass[twocolumn,showpacs,prl,aps,floatfix]{revtex4}
\usepackage{graphicx}
\usepackage{bm}
\begin{document}
\title{Soliton phase near antiferromagnetic quantum critical point in Q1D conductors}
\author{L. P. Gor'kov}
 \altaffiliation[Also at ]{L. D. Landau Institute for Theoretical Physics,
 Chernogolovka, Russia}
\author{P. D. Grigoriev$^*$}
 \email{grigorev@magnet.fsu.edu}
\affiliation{ National High Magnetic Field laboratory, Florida
State University, Tallahassee, Florida }
\date{\today }

\begin{abstract}

In frameworks of a nesting model for Q1D organic conductor at the
antiferromagnetic (SDW) quantum critical point the first-order
transition separates metallic state from the {\it soliton phase}
having the periodic domain structure. The low temperature phase
diagram also displays the 2nd-order transition line between the
soliton and the uniformly gapped SDW phases. The results agree with
the phase diagram of (TMTSF)$_2$PF$_6$ near critical pressure [T.
Vuletic et al., Eur. Phys. J. B 25, 319 (2002)]. Detection of the
2nd-order transition line is discussed. We comment on
superconductivity at low temperature.

\end{abstract}



\maketitle

We address the issue of quantum critical point (QCP) for the
itinerant antiferromagnetism (AFM) (also known as spin-density
wave state (SDW)) in Q1D compounds. As it was pointed out in
\cite{Hertz}, the standard renormalization group (RG) analysis is
not applicable for superconductivity QCP and theoretical models
for AFM with nesting features because no expansion in a form of a
Landau functional is possible near QCP at $T=0$ in these cases.
QCP with increase in pressure have been observed for SDW in the
Bechgaard salts (TMTSF)$_2$X (with X=PF$_6$ \cite{Jerome1980},
AsF$_6$ \cite{Brus}) and recently investigated in many details in
\cite{Vuletic}. According to \cite{Vuletic}, the phase diagram of
(TMTSF)$_2$PF$_6$ near its critical pressure, $p_c\sim 9.4$kbar,
indeed, has a rather complicated character. However, the
experimental fact that the transition from metallic to SDW state
is of the first order suggests that quantum fluctuations do not
play decisive role near $p_c$, and the overall phase diagram near
$p_c$ for (TMTSF)$_2$PF$_6$ can be understood already on a mean
field level.

The most unexpected finding in \cite{Vuletic} is the discovery of a
pressure interval below $p_c$, where SDW and metal phases (SDW and
superconductivity (SC) at lower $T$) coexist as parallel domains
running perpendicular to the chain direction. Such coexistence is
difficult to understand for the 1st order transition that takes
place at constant pressure \cite{Vuletic}. In what follows we
propose that such domains can be interpreted as formation of a new
phase of the soliton walls suggested first theoretically for the Q1D
electron spectrum in \cite{BGS,BGL,GL}. Domains inside the SDW
interval were also observed in \cite{LeeChaikin}. We shall see that
it is rather natural to expect SC in these domains at lower
temperature. In this presentation we restrict ourselves mainly by
the effects related to the physics of SDW.  Superconductivity
appears at lower temperature than the onset of the SDW/metal phase
transition, and we only comment on it at the end.

For the SDW description we adopt the standard Q1D model of two
open Fermi surface (FS) sheets with the energy spectrum of free
electrons in the form:
 \begin{equation}
   \label{1}
   \varepsilon (\boldsymbol{p}) = \pm v_F(p_x \mp p_F) +
   t_{\perp}(\boldsymbol{p}_{\perp}),
 \end{equation}
with dispersion in the direction perpendicular to the chains
 \begin{equation}
   \label{1t}
t_{\perp}(\boldsymbol{p}_{\perp}) =
t_b(\boldsymbol{p}_{\perp})+t'_b(\boldsymbol{p}_{\perp}).
 \end{equation}
The condition
 \begin{equation}
   \label{1tt}
\left<
t_b(\boldsymbol{p}_{\perp})\right>_{\boldsymbol{p}_{\perp}}
=\left<t'_b(\boldsymbol{p}_{\perp})\right>_{\boldsymbol{p}_{\perp}}
=0
 \end{equation}
is assumed to conserve the number of electrons. The angular
brackets mean averaging over $p_{\perp}$:
$$ <..>_{\boldsymbol{p}_{\perp}}\equiv \int (..)(b/2\pi)dp_{\perp} .
$$
The component $t_b(\boldsymbol{p}_{\perp})$ possesses the
so-called ''nesting'' feature (at $t'_b\equiv 0$) favoring
formation of the SDW gap along the two sheets of the Fermi
surface. For that one needs
 \begin{equation}
   \label{2}
   \varepsilon (\boldsymbol{p}+\boldsymbol{Q})=-\varepsilon (\boldsymbol{p}).
 \end{equation}
''Antinesting'' term, $t'_b(\boldsymbol{p}_{\perp})$, destroys the
perfect nesting condition (\ref{2}). Increase of $t'_b$ simulates
increase in applied pressure. At large enough $t'_b$ SDW
disappears.

 The model assumes a bare e-e interaction of the form
 \begin{equation}
   \label{3}
   u(\boldsymbol{Q})\boldsymbol{\sigma_1}\boldsymbol{\sigma_2},
 \end{equation}
where $u(\boldsymbol{Q})$ has a maximum at some $\boldsymbol{Q}_0$.
Interaction (\ref{3}) is renormalized after summing up the ladder
diagrams:
 \begin{equation}
   \label{5}
   U(\boldsymbol{Q})=\frac{u(\boldsymbol{Q})}{1+u(\boldsymbol{Q})
   \Pi (\boldsymbol{Q},\omega_0)}.
 \end{equation}
The polarization operator
 \begin{equation}
   \label{6}
   \Pi (\boldsymbol{Q},\omega )=T\sum_{\omega_n}\int\frac{d^2\boldsymbol{p}}{(2\pi
   )^2}
   G(\boldsymbol{p},\omega_n)G(\boldsymbol{p}-\boldsymbol{Q},\omega_n-\omega )
 \end{equation}
is proportional to the staggered susceptibility. Its value is
large near a ''nesting'' vector $\boldsymbol{Q}_0$. Most often
this vector is chosen as
 \begin{equation}
   \label{4}
\boldsymbol{Q}_0=(\pm 2p_F,\pi /b, \pi /c),
 \end{equation}
although the majority of the results below do not change, at least
qualitatively, at different optimal ''nesting'' wave vectors
$\boldsymbol{Q}_0$. For brevity, in (\ref{1t}) we shall leave out
the dependence on the third direction along the $c$-axis
\cite{Com2}.

When the denominator in (\ref{5}) becomes zero, the metal phase is
absolutely unstable with respect to a SDW(CDW) nucleation. The
instability line $T_c(t_b',\boldsymbol{Q})$ is  given by a
well-known equation:
 \begin{equation}
   \label{7}
\ln\left(\frac{T_{c0}}{T_c}\right) =\mbox{Re} \left<
\Psi\left(\frac12+\frac{i W(\boldsymbol{Q},p_{\perp})}{4\pi
T_c}\right)\right>_{p_{\perp}}-\Psi\left(\frac12\right) ,
 \end{equation}
where $T_{c0}$ is the temperature of the SDW onset at $t_b'=0$ and
$W(\boldsymbol{Q},p_{\perp})= \hbar v_F(Q_x-2k_F)+ t_{\perp}(p_y)+
t_{\perp}(p_y-Q_y)$. Eq. (\ref{7}) has been analyzed by many
authors (see, e.g. \cite{BGL,HasFuk,Montambaux,Jafarey}). At zero
temperature Eq. (\ref{7}) writes down as \cite{BGL}
 \begin{equation}
   \label{8}
\left< \ln\left(\pi T_{c0}/\gamma W(\boldsymbol{Q},p_{\perp})
\right) \right>_{p_{\perp}}=0,
 \end{equation}
where $\gamma\approx 1.781$ is the Euler constant.

With increasing pressure the value of $t_b'$ increases, and at
some critical pressure $p_c$ the SDW ordering may disappear even
at zero temperature. This point could be considered as QCP.
However, one sees that the application of renormalization group
(RG) analysis \cite{Hertz} to QCP in Q1D case would meet
difficulties because an analytic expansion of Eq. (\ref{8}) at
small $\boldsymbol{q}\equiv \boldsymbol{Q}-\boldsymbol{Q}_0$ does
not exist due to the condition (\ref{1tt}).

The integral (\ref{8}) is convergent at any finite
$\boldsymbol{Q}$, and one can find critical $t_b'$ for the general
form of $t_{\perp}(p_{\perp})$. It does not necessarily occur at
$\boldsymbol{Q}= \boldsymbol{Q}_0$, but at some
$\boldsymbol{Q}^*\equiv\boldsymbol{Q}_0+\boldsymbol{q}^*$
\cite{Comment1}. Transition into such a new SDW phase with
''shifted'' wave vector (SDW$_2$ phase) should be of the second
order, and, hence contradicts to the observed 1st-order character
of the transition between SDW and the metal phase in
\cite{Vuletic}.

It is known that for strictly 1D models of CDW (or SDW) the
character of excitations is different from the ordinary one.
Instead of electron-hole pairs with excitation energy $\approx
2\Delta $, the propagating excitations are solitons (or soliton
''kinks'') that cost a lower energy and come about together with
the nonlinear reconstruction of the underlying gaped ground state
\cite{Su,Braz}. (TMTSF)$_2$PF$_6$ has a quarter filled band
($2k_F=\pi /2a$). Unlike the Peierls state in polyacetylene,
(CH)$_x$ , commensurability effects play here no essential role
\cite{JeromeReview}, so that only the neutral spin $1/2$ solitons
exist \cite{Braz}. Below we merely borrow the results from the
extended literature  on solitons in a 1D chain (for a review and
references, see \cite{BrazKirovaReview}). Although this literature
is devoted to the physics of CDW state, all the results can be
transferred to the SDW case, practically, without changes.

The energy cost for one soliton ''kink'' on a {\it single} 1D
chain is
 \begin{equation}
   \label{10}
E_s=(2/\pi)\Delta_0.
 \end{equation}
For chain packed in a crystal a tunneling, $t_{\perp}(p_{\perp})$,
arise between the chains. Therefore, the solitons form extended
states by creating a band in the transverse direction. Instead of
independent kinks on different chains, one arrives at a soliton wall
\cite{BGS} with neutral solitons occupying two spin states with
$t_{\perp}(p_{\perp})<0$ in the band . The energy cost
$A(t_{\perp})$ of the wall referred to one chain decreases,
 \begin{equation}
   \label{12}
A(t_{\perp})=(2/\pi)\Delta_0-2\int_{t_{\perp}\leq 0}
t_{\perp}(p_{\perp}) b dp_{\perp}/2\pi ,
 \end{equation}
and may even become negative. In other words, the spontaneous
formation of soliton wall may be favorable at large enough
$t_{\perp}$.

Eq.(\ref{12}) presents our key idea for interpreting the phase
diagram of (TMTSF)$_2$PF$_6$ \cite{Vuletic}: with the pressure
increase the system first reaches a critical pressure, $p_{c1}$, at
which $A(t_{\perp})=0$, and smoothly enters into a new phase -- the
soliton phase (SP). At some higher pressure ($p_{c}$) the
first-order transition between SP and metallic phase takes place.

To describe the sequence of such transitions, one needs expressions
for the energy of SP. The problem was solved in \cite{BGL} for the
CDW ordering and for ''direct'' nesting vector
$\boldsymbol{Q}_0=(\pm 2k_F,0)$. Change from CDW to SDW needs no
explanation. Without staying on the proof, we remark that the gauge
transformation of the wave functions, as in \cite{BGL}, removes the
large ''nesting'' term $t_b(p_{\perp})$ from all expressions below.
As the result, it is only the ''antinesting'' term $t'_b(p_{\perp})$
that enters Eq. (\ref{12}): $t_{\perp}(p_{\perp})\rightarrow
t'_b(p_{\perp})$.

We now write down the expression for the linear energy density
$W_{SP}$ of the soliton phase \cite{BGL} in the limit of large
distance between the walls($T=0$):
 \begin{equation}
   \label{13}
W_{SP}=-\frac{\Delta_0^2}{2\pi \hbar v_F} + n A(t_{b}^{'})+n E_-^2
B,
 \end{equation}
where $n$ is the walls' linear density, and the $E_-^2$ term in
(\ref{13}) correspond to the exponentially decaying interaction
between the walls. Indeed, in this limit of large distances
between walls, $n$ from \cite{BGL}
 \begin{equation}
   \label{14}
n=\frac{E_+/\hbar v_F }{K\left(\sqrt{1-E_-^2/E_+^2}\right) }
\approx \frac{\Delta_0/\hbar v_F }{\ln\left( 4\Delta_0/E_-\right)
},
 \end{equation}
where $E_+\approx \Delta_0$ at $n\to 0$ and $K(r)$ is the complete
elliptic integral of the 1st kind. From (\ref{14})
$$
\label{14t} E_-\approx 4\Delta_0\exp (-\Delta_0/n\hbar v_F) .
$$
$A(t_{b}^{'})$ is given by Eq. (\ref{12}) with
$t_{\perp}(p_{\perp})= t^{'}_b(p_{\perp})$, and
 \begin{equation}
   \label{15}
B(t_{\perp})=\frac{1}{2\pi\Delta_0
}-\frac{b}{|dt^{'}_b/dp_{\perp}|_0},
 \end{equation}
where $|dt^{'}_b/dp_{\perp}|_0$ is the value of the transverse
velocity at the four point where $t^{'}_b(p_{\perp})=0$. At $B>0$
crossing the point $A(t^{'}_b)=0$ corresponds to the second-order
transition from the ''homogeneous'' SDW to the soliton walls lattice
state. Negative $B<0$ would mean an abrupt first-order phase
transition \cite{BGL}. All the results (\ref{12}-\ref{15}) are for
$T=0$.

We now return to Eq. (\ref{7}). As the absolute instability line, at
higher temperature it defines the 2nd-order transition line between
the metal and SDW(CDW) phase. At lower temperature for most
$t_{b}^{'}(p_{\perp })$ models it corresponds to some ''shifted''
$\boldsymbol{Q}^*$. As it was mentioned above, experimentally
\cite{Vuletic}, the metal-to-SDW transition at lower temperature is
of the 1st-order. At these temperatures the line defined by Eq.
(\ref{7}) has only the meaning of a ''supercooling'' line.

Theoretically, the positions of critical points $p_c$ and $p_{c1}$
depend essentially on a particular form of $t'_b(p_{\perp})$. The
function $t'_b(p_{\perp})$ in (TMTSF)$_2$PF$_6$ is unknown.
Therefore, for qualitative analysis we will apply the expressions
(\ref{14}),(\ref{15}) also in the regime of dense soliton phase. We
compare two examples: first, the oversimplified model, in which
$t'_b(p_{\perp})$ has the periodic step-like shape:
 \begin{equation}
   \label{16}
t^{'}_b(p_{\perp})=\biggl\{
\begin{array}{c}
2t^{'}_b,\  0<p_{\perp}<\pi /b\\
-2t^{'}_b,\  \pi /b <p_{\perp}<2\pi /b \;,
\end{array}
 \end{equation}
and the second one with the tight-binding dispersion of the form
\begin{equation}
t'_b(p_{\perp})=2t_b'\cos(2p_{\perp}b)
   \label{9}
\end{equation}

We calculated the energy of the soliton phase for both dispersion
functions. At the calculation the optimal value of $E_-$ (or $n$)
must be found to minimize the energy (\ref{13}) of the soliton
phase. For the first dispersion (\ref{16}) the domain wall energy
$A(t'_b)$ becomes negative at $t'_b\geq t_{c1}\approx 0.32\Delta_0$.
This point corresponds to the 2nd-order transition from
''homogeneous'' SDW phase into the soliton phase. Then we compared
the soliton phase energy $W_{SP}$ with the energy of metal phase
 \begin{equation}
   \label{17}
W_n=-(1/\pi\hbar v_F)\left< [t'_b(p_{\perp})]^2\right>_{p_{\perp}}
 \end{equation}
and found the point $t_{c}\approx 0.54\Delta_0$ where these energies
become equal to each other. This point corresponds to the
first-order phase transition from SP to the normal-metal phase. We
see that the interval $t_{c}-t_{c1}\approx 0.4t_{c}$ of the
existence of SP for this particular model is rather large. One can
show that for the model described by Eqs. (\ref{12})-(\ref{15}) the
function (\ref{16}) correspond to the largest possible interval of
SP.

For the second dispersion relation (\ref{9}) the similar
calculation gives equal values for $t_{c}$ and $t_{c1}$:
$t_{c}=t_{c1}=\Delta_0/2$, i.e. in this case an interval for
soliton phase is absent. This once more shows the high sensitivity
of the SDW phase diagram to the particular form of
$t^{'}_b(p_{\perp})$.

In experiment \cite{Vuletic} the difference $t_{c}-t_{c1}\approx
0.1t_{c}$ is intermediate between the cases of Eq. (\ref{16}) and
Eq. (\ref{9}), but closer to the second case of tight-binding
dispersion relation (\ref{9}) than to the step-like dispersion
(\ref{16}). The ratio $(t_{c}-t_{c1})/t_{c}\approx 0.1$ observed in
experiment \cite{Vuletic} can be easily fitted by an appropriate
choice of the model function $t^{'}_b(p_{\perp})$ (for example, by
adding a fourth harmonic to the tight-binding dispersion (\ref{9}).

In Fig. 1 we show schematically the SDW phase diagram near QCP. The
slope of the second-order transition line near $T=0$ comes about
because at $T\neq 0$ the second term in $A(t_{\perp})$ decreases to
account for thermal excitations inside the domain wall from the
occupied part of band $t^{'}_b(p_{\perp})<0$. The temperature slope,
of course, again depends on the $t^{'}_b(p_{\perp})$-dependence.

In \cite{Vuletic} the phase diagram of (TMTSF)$_2$PF$_6$ has been
studied through the resistance measurements along the chain
direction. The 1st-order character of the transition between metal
and SDW states follows from hysteretic effects, while the appearance
of domains was derived from the changes in resistivity behavior near
$p_c$. We suggest that the line of the 2nd-order transition between
the insulating ''homogeneous'' and soliton-wall phases could be
detected by measuring a sharp anisotropy of conductivity near this
transition \cite{GL}. Indeed, according to \cite{GL}, at low $n$
(large distance between the domain walls) conductivity along the
chain direction remains very low because of exponentially small
overlap between the electron wave functions inside each wall. As for
transverse conductivity, it would increase linearly with $n$.

One also sees that the value of $t'_b$ itself is not the only
parameter that may determine the dependence on $t'_b$ (pressure).
Indeed, the expression for $B(t_{\perp})$ (\ref{15}) essentially
depend on $|dt^{'}_b/dp_{\perp}|_0$ as well. The $B$-term is not
analytic at $t'_b\to 0$; change in its sign will immediately change
the whole physics \cite{BGL}.

\begin{figure*}
\begin{center}
\includegraphics{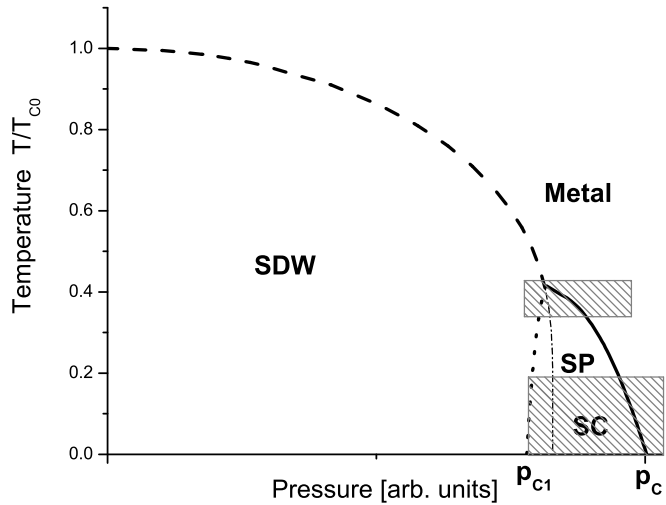}
\caption{\label{fig1} A schematic view of the SDW phase diagram. The
solid line signify the first-order transition between the metal and
soliton phases. The dot line stands for the second-order transition
between uniformly gapped SDW phase and the SP. The dash line
represents the line of absolute instability of metal phase toward
the formation of SDW. At higher temperatures this line represents
the second-order transition line between the SDW and metal phases.
At lower temperature (dash-dot line) it has only the meaning of
supercooling line. The lower filled bars shows the region where
superconductivity appears. We do not calculate the superconducting
critical temperature. The upper filled bar shows the
high-temperature region of soliton phase where substantial
quantitative deviations from formulas (\ref{12})-(\ref{15}) may
start due to the thermal excitations in the SDW and soliton phases.
Detailed shape of all lines is sensitive to the choice of the
''antinesting'' term $t^{'}_b(p_{\perp})$ (see text).}
\end{center}
\end{figure*}

In Fig. 1 the superconductivity, seen experimentally in
\cite{Vuletic} near $p_c$ and in the metallic phase at lower
temperatures ($T_c\sim 1K$), is not shown in order not to overcrowd
the phase diagram. Meanwhile, appearance of superconductivity is
inevitable in frameworks of nesting mean-field models. Indeed, the
logarithmic singularity that enters the expression for SDW
polarization operator (\ref{6}), would appear also for the Cooper
channel \cite{BGD}. At higher temperature the SDW phase prevails
because the bare interaction (\ref{3}) was chosen to be larger for
the transverse nesting component $\boldsymbol{Q}_{\perp}$. However,
with increase of ''antinesting'' term $t'_b$ (pressure) the SDW
transition becomes suppressed and SC which is not sensitive at all
to any of the $t'_b$-terms, finally becomes more favorable. To some
surprise, $T_{SC}$ for the SC transition is experimentally not
sensitive to the formation of the SP. We explain this fact
qualitatively by assuming that the SC coherence length,
$\xi_0\sim\hbar v_F/T_{SC}$, is considerably larger than the
expected periodicity for the SP of the order of $\hbar
v_F/\Delta_{0} \sim \hbar v_F/T_{0}$ resulting in considerable
Josephson coupling between the soliton walls. However, the problem
needs some further analysis.

 To summarize, we have shown that in the vicinity of QCP in
 Q1D materials, such as the Bechgaard salts, one
meets with a new soliton wall phase. As the pressure increases from
$p=0$, one first crosses the line $p_{c1}$ of the second-order
transition and enters into the periodic soliton structure with a
characteristic pressure-dependent periodicity of the order of $\hbar
v_F/\Delta_0\sim 10^3$\AA . At higher pressure, $p_c$, by the
first-order phase transition the system goes over into metallic
state. Details of the phase diagram strongly depend on material
parameters, in particular on the form of the ''antinesting''
tunneling in the electron dispersion. Coexistence of SDW and
superconductivity inside soliton walls has not been investigated.
Qualitatively one would expect the survival of superconductivity in
the ''metallic sheets'' (domain walls) perpendicular to the chain
direction \cite{GL}.

\medskip The authors acknowledge useful discussions with S.A.
Brazovskii. The work was supported by NHMFL through the NSF
Cooperative agreement No. DMR-0084173 and the State of Florida, and
(PG) in part, by DOE Grant DE-FG03-03NA00066.


\begin{thebibliography}{99}
\bibitem{Com2} In the most Q1D organic compounds the electron dispersion in
$c$-direction is very weak, and its ''antinesting'' part (that the
only influences the phase diagram) is negligible.
\bibitem{Comment1} There exists the extended literature
devoted to the two SDW states in the Q1D compounds (see, e.g.,
\cite{HasFuk,Montambaux}). It dates back to the numerical
calculations in \cite{HasFuk}. These calculations contains some
wrong results that we correct in a separate publication.
\end{thebibliography}
\end{document}